\documentclass[prl,reprint,showpacs,floatfix,amsmath,amssymb,superscriptaddress]{revtex4-1}	
\usepackage{graphicx}	
\usepackage{bm}% bold math	
\usepackage{float}	
\usepackage{wasysym}	
\usepackage[dvips]{color}	
\usepackage{hyperref}	
\usepackage{hypernat}	
\usepackage{wrapfig}
\definecolor{grey}{RGB}{255,200,180}	
\usepackage{sidecap}	
\def\etal{{\em{et al}}}

\def\K{\mathbf{K}}	
	
\begin{document}	
\title{Metal-Insulator-Transition in a Weakly interacting Disordered Electron System} 	
%Authors
\author{C. E. Ekuma}
\altaffiliation{Electronic address: cekuma1@lsu.edu}
\affiliation{Department of Physics \& Astronomy, Louisiana State University,
Baton Rouge, Louisiana 70803, USA}
\affiliation{Center for Computation and Technology, Louisiana State University, Baton Rouge, Louisiana 70803, USA}

\author{S.-X. Yang}
\affiliation{Department of Physics \& Astronomy, Louisiana State University,
Baton Rouge, Louisiana 70803, USA}
\affiliation{Center for Computation and Technology, Louisiana State University, Baton Rouge, Louisiana 70803, USA}

\author{H. Terletska}
\affiliation{Department of Physics \& Astronomy, Louisiana State University,
Baton Rouge, Louisiana 70803, USA}
\affiliation{Center for Computation and Technology, Louisiana State University, Baton Rouge, Louisiana 70803, USA}

\author{K.-M. Tam}
\affiliation{Department of Physics \& Astronomy, Louisiana State University,
Baton Rouge, Louisiana 70803, USA}
\affiliation{Center for Computation and Technology, Louisiana State University, Baton Rouge, Louisiana 70803, USA}

\author{N. S. Vidhyadhiraja}
\affiliation{Theoretical Sciences Unit, Jawaharlal Nehru Center for Advanced Scientific Research, Bangalore, 560064, India}

\author{J. Moreno}
\affiliation{Department of Physics \& Astronomy, Louisiana State University,
Baton Rouge, Louisiana 70803, USA}
\affiliation{Center for Computation and Technology, Louisiana State University, Baton Rouge, Louisiana 70803, USA}

\author{M. Jarrell}
\altaffiliation{Electronic address: jarrellphysics@gmail.com}
\affiliation{Department of Physics \& Astronomy, Louisiana State University,
Baton Rouge, Louisiana 70803, USA}
\affiliation{Center for Computation and Technology, Louisiana State University, Baton Rouge, Louisiana 70803, USA}
%=======================================================================================================	
\begin{abstract}	
\noindent 	
The interplay of interactions and disorder is studied using the Anderson-Hubbard model within the typical 
medium dynamical cluster approximation. Treating the interacting, non-local cluster self-energy 
($\Sigma_c[{\cal \tilde{G}}](i,j\neq i)$) up to second order in the perturbation expansion of 
interactions, $U^2$, with a systematic incorporation of non-local spatial correlations and diagonal 
disorder, we explore the initial effects of electron interactions ($U$) in three dimensions. We find that 
the critical disorder strength ($W_c^U$), required to localize all states, increases with increasing $U$; 
implying that the metallic phase is stabilized by interactions. Using our results, we predict a soft 
pseudogap at the intermediate $W$ close to $W_c^U$ and demonstrate that the mobility edge 
($\omega_\epsilon$) is preserved as long as the chemical potential, $\mu$, is at or beyond the mobility 
edge energy.
\end{abstract}	
%=======================================================================================================	
	
%=======================================================================================================	
\maketitle 	
\textit{Introduction}.--	
The metal-insulator transition (MIT) driven by random impurity has been an important topic in physics since the pioneer
work by Anderson~\cite{Anderson}. A significant advance in the MIT theory
is achieved by studying it in the context of critical phenomena. Concepts from scaling, renormalization group (RG), 
and random matrix theory are used to understand the mechanism of 
localization at different dimensions for different symmetry classes~\cite{gang4,Wegner1975,Kramer,Mirlin-RMP}. 
It has been demonstrated that an infinitesimal amount of disorder can lead to localization for the models in the 
orthogonal class at lower (one and two) dimensions, whereas there is a MIT for three dimensions (3D)~\cite{gang4}. In 3D, 
a sharp mobility edge separating localized and delocalized states develop as disorder strength increases~\cite{Mott}.

While the MIT of non-interacting systems by now is fairly well understood~\cite{Kramer,Mirlin-RMP,50years},
earlier studies suggested that interaction could play an important role in the MIT~\cite{Belitz-Kirkpatrick-1994}.
Over the last few decades, experimental works ranging from doped semiconductors~\cite{Mott,Anthony2010,Kravchenko-2d}, 
perovskite compounds~\cite{PhysRevB.76.165128,PhysRevB.71.125104,Sahu2011523,PhysRevB.74.104419,Raychaudhuri}),
to cold atoms in optical lattices\cite{sanchez2010disordered,1751-8121-45-14-143001,PhysRevLett.102.055301,Kondov13}
have highlighted the importance of the interplay of disorder 
($W$)~\cite{Anderson,gang4,Kramer,Mirlin-RMP} and interactions ($U$)~\cite{Mott}.

At the Fermi level, Altshuler-Aronov~\cite{Altshuler} showed that interactions can induce a square-root and logarithmic 
singularity in two and three dimensions, respectively, while Efros-Shklovskii demonstrated 
the Coulomb gap~\cite{Efros1975}. 
Field theory perturbative RG method and diagrammatic theory which go
beyond the Hartree-Fock approximations have suggested a metallic state for two 
dimensions~\cite{Finkel1983,Castellani1984}. 
The recent RG work by Finkelstein and co-workers has further 
indicated the possibility of a MIT for a model with degenerate valleys~\cite{Punnoose14102005}, the 
validity of which was confirmed through experiments in Si-MOSFETs~\cite{Punnoose14102005,Anissimova}.

In this letter, we focus on the system with weak local interactions on disorder systems in 3D.
Our approach is an extension of the recently developed typical
medium dynamical cluster approximation (TMDCA), which has shown to be highly successful
in describing the Anderson localization transition (ALT) for the non-interacting systems~\cite{PhysRevB.89.081107}.
The typical medium approaches assume that the typical density of states (TDoS), when appropriately defined,
acts as the ``proper'' order parameter for the ALT. Such an assumption is well justified
not only for the non-interacting case~\cite{Vlad2003,PhysRevB.89.081107,Vollhardt_PDF} but also in the
presence of interactions, as shown experimentally\cite{Anthony2010,Wei2013}. The typical
medium theory (TMT) of Dobrosavljevi\'{c} \etal~\cite{Vlad2003} is a special case of the TMDCA when the
cluster size $N_c=1$. Even though the TMT cannot include weak localization effects due
to coherent backscattering, it still does qualitatively predict a disorder-driven ALT, and hence
incorporates 'strong localization' effects. The TMDCA incorporates non-local effects via systematic finite cluster
increment and achieves almost perfect agreement with numerical exact calculations. 
%This is in sharp contrast to the local Hatree-Fock approximation where the critical disorder and the associated exponents
%are quite a bit off from accurate numerical values~\cite{Amini-2014}. 
The extension of the 
TMT to finite interactions show that interactions screen the
disorder\cite{PhysRevLett.110.066401,PhysRevLett.102.156402,Vollhardt}.
In this letter, we show that such a conclusion is robust in the
thermodynamic limit through increasing cluster size calculations.

While there have been significant efforts to understand the combined effect of 	
disorder and interactions on the local density of states close to the Fermi level,	
the band edges have received scant attention. Specifically, the effect	
of weak interactions on the mobility edge has not been discussed thus far. 
We are particularly interested in the evolution of the mobility edge under the influence 
of the Hubbard interaction for spin$-1/2$ system. The transition between 
metal--the Fermi liquid phase; and insulator--the Anderson localized phase is 
discussed, whereas the possibility of the Mott insulator is excluded in this study, as only 
short range, weak interactions will be considered. 
%Recent studies on the detail of the wavefunction, such as its 
%multifractal behavior is very interesting by itself especaially for $1/r$ interaction~\cite{Amini-2014,PhysRevLett.110.066401}, 
%though not directly relevant to our present study.
%Do we actually need this part?

The main result of this letter is that for $\mu<\omega_\epsilon$, arbitrary small 
interactions lead to the masking of the sharp mobility edge that separates localized	
and extended states in the non-interacting regime below the critical disorder strength 
$W_c^{U=0}$. Thus, interactions can radically modify the spectrum 
of a non-interacting system at the band edges, i.e., in the `localized band'. 
However, when the chemical potential ($\mu$) is at or above the mobility 
edge energy (i.e., $\mu \geqq \omega_\epsilon$), the well-defined localization 
edge is restored. Nevertheless, 
% with an important exception. 
unlike the non-interacting systems where the TDoS just shifts 
rigidly as one scans through $\mu$, in the presence of interactions, there is a 
non-trivial decrease of the TDoS vis-\`{a}-vis the change in the filling. 
Further, we find a soft-pseudogap at intermediate $W$ just below $W_c^{U}$.

%======================================================================================================= 	
\textit{Method}.--	
The Anderson-Hubbard model (AHM) is a model for studying the interplay 	
between electron--electron interactions and disorder. The Hamiltonian for this model is	
\begin{equation}	
  \label{eqn:model}	
H=-\sum_{\langle i j \rangle \sigma}t_{ij}(c_{i \sigma}^{\dagger}c^{\phantom{\dagger}}_{j \sigma}+h.c.)+\sum_{i \sigma} (V_i-\mu)n_{i \sigma} + 	
U\sum_{i} n_{i \uparrow} n_{i \downarrow}.	
\end{equation}	
%The first term is the energy operator due to hopping of electrons on the lattice. The
The first term describes the hopping of elect					rons on the lattice, 
$c_{i}^\dagger$($c^{\phantom{\dagger}}_{i}$) is the creation (annihilation) operator of an electron on site $i$ 
with spin $\sigma$,
$n_{i} = c_{i}^\dagger c^{\phantom{\dagger}}_{i}$ is the number operator, $t_{ij}=t$ is the hopping matrix 	
element between nearest-neighbor sites. The second term represents 	
the disorder part which is modeled by a local potential $V_i$ randomly distributed according 	
to a probability distribution $P(V_i)$, $\mu$ is the chemical potential. 
%The last term 	depicts a two-body term which is the energy cost, $U > 0$ due to Coulomb repulsion for double 	
%occupancy of a site.
The last term describes the Coulomb repulsion between two electrons occupying site $i$.
We set $4t = 1$ as the energy unit and use a ``box'' distribution with 	
$P(V_i)=\frac{1}{2W}\Theta(W-|V_i|)$, where $\Theta(x)$ is the Heaviside step function. We use the 	
short-hand notation: $\langle...\rangle=\int dV_i P(V_i) (...)$ for disorder averaging. 	
%The simulation now contains three independent energy scales: the disorder strength $W/4t$, 	
%the interaction strength $U/4t$, and the inverse temperature $4t\beta$, but the results 	
%reported here are for zero temperature ($T=0$). 	
	
Our focus is on the single-particle Green function and the associated density of states.	
To obtain these for the AHM (\ref{eqn:model}), we modify the TMDCA to treat 
both disorder and interactions. Here, an initial guess for the 
hybridization function ($\Gamma (\K,\omega) \equiv \Im 10^{-2}$) is used 
%(if there is a priori knowledge of the cluster self-energy 
%and Green function $G^c(\K,\omega)$) else, $\Gamma (\K,\omega) \equiv 0$ may serve as the starting point 
to form the cluster-excluded Green function ${\cal G}(\K,\omega)=(\omega-\Gamma(\K,\omega)-\bar{\epsilon}_\K +\mu)^{-1}$, 	
where $\bar{\epsilon}_\K$ is the coarse-grained bare dispersion. ${\cal G}(\K,\omega)$ is then 	
Fourier transformed to form the real space Green function, 	
${\cal G}_{n,m} = \sum_{\K} {\cal G}(\K)\exp(i \K\cdot(\mathbf{R}_n-\mathbf{R}_m))$ 	
and then for a given disorder configuration $\hat{V}$, we may calculate the cluster Green 	
function $G^c(\hat{V})=({\cal G}^{-1}-\hat{V})^{-1}$. 	

\begin{wrapfigure}{l}{0.2\textwidth}
\center{\includegraphics[trim = 0mm 5mm 5mm 15mm,width=.2\textwidth]{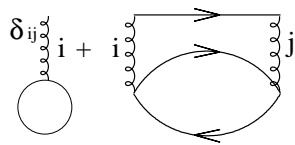}}
%\begin{SCfigure}	
%FIG. SOPT	
%\centering	
 %\includegraphics[trim = 0mm 5mm 0mm 8mm,scale=0.5]{SOPT_SE_Diagrams.xfig.eps}	
\caption{The first and second-order diagrams of the interacting self-energy between sites $i$ and $j$.}	
\label{Fig:sopt} 	
%\end{SCfigure}	
\end{wrapfigure}

Utilizing $G^c(\hat{V})$, we then calculate the Hartree-corrected cluster Green function 	
${\cal \tilde{G}}^{-1}_c(\hat{V},U) = G^c(\hat{V})^{-1} + \epsilon_d(U)$ (where $\epsilon_d(U)=\tilde{\mu} -U\tilde{n_i}/2$ 	
and $\tilde{n_i}$ $=$ $-1/\pi \int_{-\infty}^0  \Im {\cal \tilde{G}}_c(i,i,\omega) d\omega$ is the 	
site occupancy at zero temperature, $T=0$). 
Both ${\cal \tilde{G}}$ and $\tilde{n_i}$ are converged and then used to compute 
the second-order diagram shown in Fig.~\ref{Fig:sopt}. 
We note that $\tilde{n_i}$ obtained at ${\cal \tilde{G}}_c$-level is numerically 
the same as using the full Green function since $\tilde{n_i}$ is self-consistent at 
the TMDCA level. This also enables the incorporation of crossing diagrams 
(for $N_c>1$) from both disorder and interactions at equal footing and it is 
computationally cheaper (for $N_c=1$, it is $\sim$ 8 times cheaper), 
enabling simulation of large systems.

Here, we choose the chemical potential $\tilde{\mu}=\mu + U/2$ to enable simulations both at and away 
from half-filling. Thus, the full self-energy due to interactions is then
$\Sigma^{Int}_c(i,j,\omega) = \Sigma^{H}_c[{\cal \tilde{G}}]+\Sigma^{(SOPT)}_c[{\cal \tilde{G}}]$, 	
where the first term is the static Hartree correction and the second term is the non-local second-order	
perturbation theory (SOPT) contribution. We note that 	
the computational cost grows exponentially with each order of the perturbation series making it numerically 	
prohibitive to include more diagrams. However, since our focus is on the weak 	
interaction regime $U/4t \ll 1$, we expect that higher order diagrams are suppressed by at least $\sim U^3$. 

We have carried out extensive benchmarking of the 	
TMDCA-SOPT cluster solver against numerically exact quantum Monte-Carlo calculations~\cite{PhysRevB.72.035122,Rubtsov2004,
RevModPhys.83.349,0295-5075-82-5-57003,Fuchs2011,PhysRevLett.56.2521,PhysRevB.76.205120,681704,Gebhard2003} within the dynamical cluster 	
approximation (DCA) framework. For weak interactions and essentially all disorder strengths, the corrections due 	
to perturbation orders higher than the second are found to be negligible (for details, see Supplemental Material 	
(SM)~\cite{TMDCA-SOPT_Supp}).   	

%=======================================================================================================

%=======================================================================================================	
%\begin{figure}[htb]	
%%FIG. SOPT	
% \includegraphics[trim = 5mm 5mm 5mm 8mm,scale=0.5]{SOPT_SE_Diagrams.xfig.eps}	
%\caption{The first and second-order diagrams of the interacting self-energy between sites $i$ and $j$.} 	
%\label{Fig:sopt} 	
%\end{figure}		
	
For a given interaction strength $U$ and randomly chosen disorder configuration $V$, we calculate 
the fully dressed cluster Green function $\tilde G^c(\hat{V},U)=({\cal G}^{-1}-\hat{V}-\Sigma^{Int}(U)+U/2)^{-1}$.	
With $\tilde G^c(\K,\omega,V,U)$, we calculate the typical density of states as
\begin{equation}
\rho_{typ}^c(\K,\omega)= \exp\left(\frac{1}{N_c} \sum_{i=1}^{N_c} \left\langle \ln \rho_{i}^c 
(\omega,V)\right\rangle\right) \left\langle \frac{\rho^c(\K,\omega,V)}{\frac{1}{N_c} \sum_{i} \rho_{i}^c (\omega,V)} \right\rangle
\end{equation}
following the prescriptions of Ref.~\cite{PhysRevB.89.081107}, 
which avoids self-averaging. The disorder and interaction averaged typical cluster Green function is 
obtained using the Hilbert transform 	
$ G_{typ}^c(\K,\omega)=\int d \omega' \displaystyle \rho_{typ}^c(\K,\omega')/(\omega - \omega')$.	
We close the self-consistency loop by calculating the coarse-grained cluster Green function	
of the lattice 	
\begin{equation}	
\overline{G} (\K,\omega)  	
= \int \displaystyle \frac{N^c_0(\K,\epsilon) d\epsilon}{(G^c_{typ} (\K,\omega))^{-1} + \Gamma (\K,\omega) - 	
 \epsilon + \overline{\epsilon}(\K)},	
\end{equation}	
where $N^c_0(\K,\epsilon)$ is the bare partial density of states.	
	
%=======================================================================================================	
\textit{Results and Discussion}-- 	
%=======================================================================================================	
\begin{figure}[b]	
%FIG. 1	
 \includegraphics[trim = 0mm 0mm 0mm 0mm,width=1.0\columnwidth,keepaspectratio,clip=true]{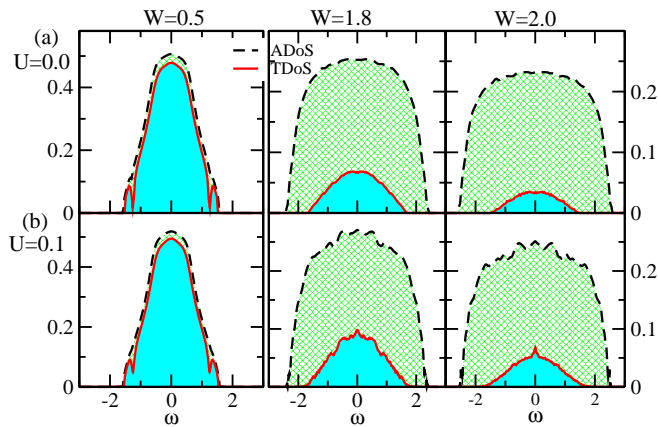}	
\caption{(Color online). Evolution of the ADoS and TDoS at various $W$ at $U=0$ 
(\textbf{a}) and $U = 0.1$ (\textbf{b}) for the TMDCA-SOPT with $N_c=38$ for the half-filled Anderson-Hubbard model (AHM). 
}
\label{Fig1:tdos_ados} 	
\end{figure}	
%=======================================================================================================	
We start the analysis of our results by comparing the algebraic 	
(or average) density of states (ADoS) (obtained from the DCA, 	
where the algebraic averaging is utilized in the self-consistency) and the typical density 	
of states (TDoS) (obtained from the TMDCA-SOPT, where the self-consistency environment is defined 	
by a typical medium) for a finite cluster $N_c=38$ at various disorder strengths for $U = 0.0$ 	
and $0.1$ at half-filling (Figs.~\ref{Fig1:tdos_ados}(a) and (b)). 	
	
At weak disorder, $W\sim0.5$, the TDoS resembles the ADoS. However, for larger $W$, 	
comparing the $U=0.0$ results (Fig.~\ref{Fig1:tdos_ados}(a)) with those of $U=0.1$ (Fig.~\ref{Fig1:tdos_ados}(b)), a noticeable 	
renormalization of the spectrum is observed.
There is a gradual suppression of the TDoS as the disorder strength is increased 
for both $U=0.0$ and $0.1$. The TDoS at $\omega=0$ is noticeably larger when	
the $U$ is finite. This indicates a delocalizing effect of interactions 
which is consistent with a real space renormalization group study~\cite{Ma-1982} and has been interpreted as a screening	
of the disorder~\cite{PhysRevLett.91.066603,PhysRevLett.110.066401}.
%, due to increased virtual processes.	
For a given disorder strength, the band edges at half-filling for the interacting case appear to be identical to that of the	
$U=0$ spectrum. This seems to imply that the mobility edge is preserved when $U$ is turned on.	
However, this is not the case, and this becomes clear upon examining the tails of the density of states.	
%=======================================================================================================	
\begin{figure}[htb!]	
%FIG. 3	
 \includegraphics[trim = 0mm 0mm 0mm 0mm,width=1.0\columnwidth,keepaspectratio,clip=true]{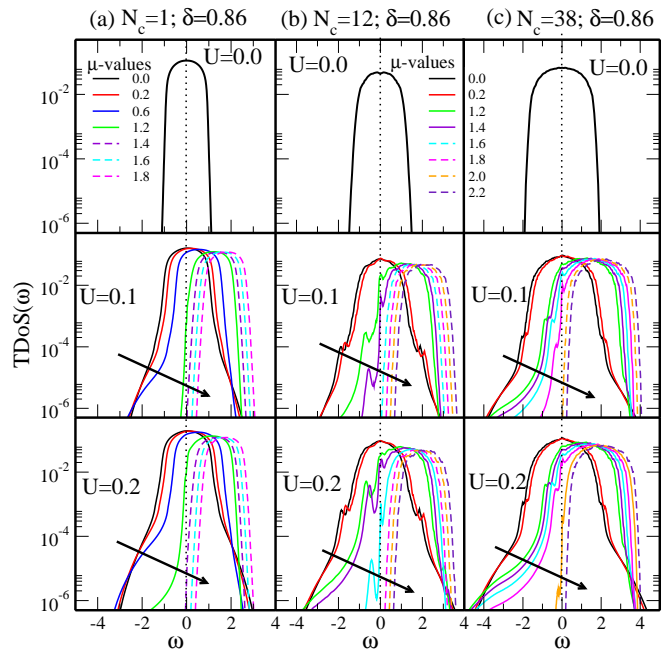}	
\caption{(Color online). The evolution of the TDoS of the AHM for 	
increasing $U$-values for the TMT ($N_c=1$, (\textbf{a})) and finite clusters 
($N_c=12$ (\textbf{b}) and 38 (\textbf{c})) at fixed $\delta=W/W_c^U=0.86$ on a log-linear plot for increasing $\mu$-values. 
For $U=0.0$, we show the plot for $\mu=0$ only, since changing $\mu$ only involves 
a rigid shift of the TDoS. For $U>0$, notice the systematic disappearance of the exponential tails 
(indicated by arrow) and the non-trivial decrease of the TDoS for the finite $U$
(unlike the rigid shift in $U=0$) as one approaches the mobility edge energy.} 	
\label{Fig:TDoS_log-linear}  	
\end{figure}	
%=======================================================================================================	
	
To explore the effect of weak interactions on the localization edge of a disordered electron system, 	
we show in Fig.~\ref{Fig:TDoS_log-linear}, the evolution of the TDoS with $\delta=W/W_c^U$ for various values of $U$ 	
on a linear log plot at various $\mu$. Clearly for $U=0$, a sharp, well-defined mobility edge is observed
(see also\ Fig.\ref{Fig1:tdos_ados}(a)). However, even for a very small 
$U=0.1$ ($1/30$ of the bandwidth), and for both the TMT and TMDCA-SOPT, the sharp localization 
edge is replaced by an exponential tail, when $\mu<\omega_\epsilon$. Hence, the incorporation 
of Coulomb interactions in the presence of disorder for $\mu<\omega_\epsilon$ leads to long band tails 
that are exponentially decaying. This fingerprint can be understood from a Fermi liquid perspective. 

If we inject an electron into a Fermi liquid with an energy $\omega$ above the Fermi energy, 
then, we expect the particle to experience an inelastic scattering, due to $U$ which is proportional 
to $\omega^2$.  One factor of $\omega$ is due to energy conservation and the other to momentum 
conservation with both constrained by the Pauli principle. I.e., the inelastic scattering 
vanishes as $\omega \rightarrow 0$.  However, if we apply the same logic to an interacting disordered system, 
then, we might expect the edge of the TDoS to be smeared out by these inelastic scattering 
processes, whenever the edge energy is above the Fermi energy, but become sharp as the edge, approaches it. 
Though, some argue that this reasoning fails for a disordered system, especially for a 
strongly disordered system since a well-defined quasiparticle no longer exists~\cite{Golubev1998164}. 
As a consequence, the concept of a mobility edge would not hold and the TDoS should have pronounced 
exponential ``tails'' even when the Fermi energy approaches the top or bottom of the TDoS bands. As it is evident from 
Fig.~\ref{Fig:TDoS_log-linear}, the sharp mobility edge is restored as the mobility edge 
energy is approached in tandem with the Fermi liquid description.

%=======================================================================================================	
\begin{figure}[t]	
%FIG. 2	
\includegraphics[trim = 0mm 0mm 0mm 0mm,width=1\columnwidth,clip=true]{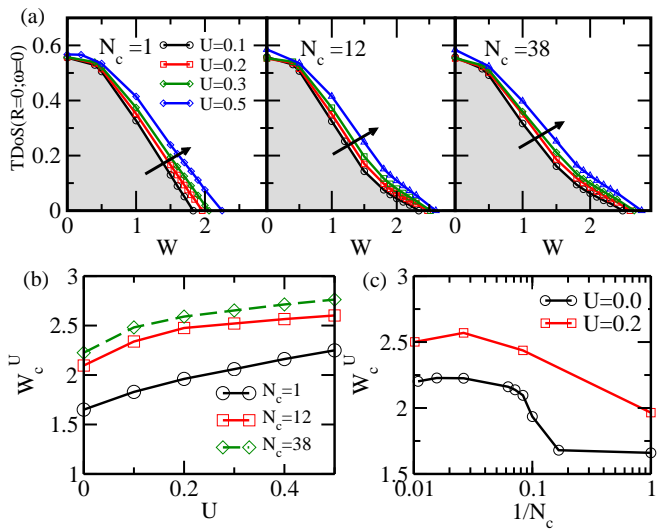}	
\caption{(Color online). (\textbf{a}) The evolution of the TDoS (at $\omega=0$) as a function 
of the disorder strength $W$ for various interactions for $N_c=$ 1, 12, and 38 at half-filling. 
The integral $\int \Im \Gamma(K,\omega)\,dK\,d\omega$ vanishes at 
the same W$_c$ as the TDoS for a given $U$ (not shown), signifying that the absence of the hybridization paths leads to the 
vanishing of the TDoS. As indicated by the arrow, increasing $U$ pushes W$_c$ to larger values.  
(\textbf{b}) The interaction dependence of the critical disorder $W^U_c$ 
for different cluster sizes $N_c =1$, 12, and 38 	
of the AHM at half-filling. The unit is fixed by setting $4t=1$. The plot is generally in 	
agreement with the results of Ref.~\cite{PhysRevLett.94.056404}. %as such, our method is valid within the $U$-values considered.
(\textbf{c}) The $W_c^{U}$ vs $1/N_c$ on a semi-log plot at $U=0.0$ and $U=0.2$ for the half-filled AHM. 
Note the systematic and fast convergence of W$_c^U$ with cluster size for both cases.}
\label{Fig2:tdos_ados_center}  	
\end{figure}	

%=======================================================================================================	
The smearing of the TDoS edge can further be inferred from the convolutions found in the second 
order (and higher) diagrams (cf. Fig.~\ref{Fig:sopt}), which will mix states above and below the 
non-interacting localization edge. Consider two such states: one 
localized and the other extended, which are now degenerate due to this mixing. 
Since these states hybridize with each other, both states will become extended~\cite{Mott}. 
%However, such scenario no more exists when the chemical potential 
%is moved outside the mobility energy (cf. Fig.~\ref{Fig:TDoS_log-linear}).

%=======================================================================================================	
\begin{figure}[t]	
%FIG. 5
\includegraphics[trim = 0mm 0mm 0mm 0mm,width=1.\columnwidth,keepaspectratio,clip=true]{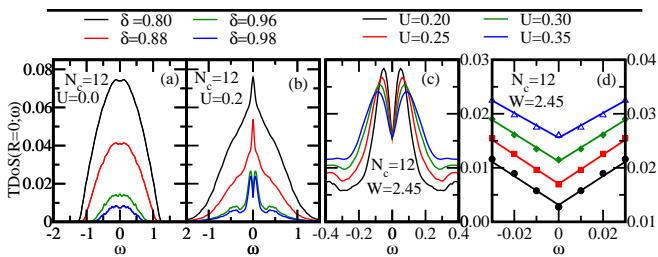}	
\caption{(Color online). The TDoS vs energy ($\omega$) for $N_c=12$ ($U=0.0$) (\textbf{a}) and 
$N_c=12$ ($U=0.2$) (\textbf{b}) at various $\delta=W/W_c^U$ showing the formation of a pseudogap at intermediate 
$W$ just before $W_c^{U=0.2}$, which is absent when $U=0$. (\textbf{c}) Shows the TDoS vs $\omega$ at 
a fixed $W$ (close to $W_c^{U=0.2}=2.50$) for various $U$. Note, the data has been scaled with $U$. 
(\textbf{d}) Same data as in Fig.~\ref{Fig2:pseudo_gap}(c) showing the linear dependence of the pseudogap on $\omega$.}
\label{Fig2:pseudo_gap}  	
\end{figure}	

%=======================================================================================================	
Next, we explore the effect of interactions on the half-filled, disorder-driven localization transition.	
We show in Fig.~\ref{Fig2:tdos_ados_center} the evolution of the TDoS at the band center, $\omega=0$, 
for various cluster sizes. The integrated escape rate ($\int \Im \Gamma(K,\omega)$$dKd\omega$) 
(not shown) characterizes the rate of diffusion of electrons between the impurity/cluster and the typical medium. 
The vanishing of the hybridization paths leads to a localization transition. The TDoS vanishes at the same 
value of $W_c^{U}$ as the integrated escape rate.

Figure~\ref{Fig2:tdos_ados_center}(a) shows that an increase in $U$ from 0.1 to 0.5 leads to a concomitant 	
increase in $W_c^U$. One can say loosely that, the zero-temperature effect of correlations is an 	
effective reduction in the disorder strength~\cite{PhysRevLett.91.066603,Vollhardt}, leading to the 	
increase in $W_c$ as indicated by the arrow. For the 	
TMT ($N_c=1$), the $W_c^U$ increases as 1.83, 1.96, 2.06, 2.16, and 2.25 for $U=$ 0.1 -- 0.5, 
while for the TMDCA ($N_c=12$), $W_c^U$ increases as 2.34, 2.48, 2.52, 2.57, and 2.60, and for 
for the TMDCA ($N_c=38$) as 2.48, 2.59, 2.65, 2.71, and 2.76 for $U=$ 0.1 -- 0.5.
%We note that $W_c^U$ increases more quickly with $U$ as one goes from single-site ($N_c=1$) to 	
%finite clusters ($N_c=12$ and 38). This is likely due to the effect of a finite $U$ on the coherent	
%backscattering, which is absent for $N_c=1$ and is systematically incorporated as $N_c$ increases.  

In Figure~\ref{Fig2:tdos_ados_center}(b), we show the interaction $U$ dependence of 
the critical disorder strength $W_c^U$ for N${}_c=1$, 
$12$, and $38$ for the half-filled AHM. For each of the $N_c$, we obtain a correlated metal below the lines, 
and above we have the gapless Anderson-Mott insulator. The trend in both the single site and finite cluster 
are alike (i.e., $W_c^U$ increases with increasing $U$) except for the difference in $W_c^U$. 	
The almost linear trend observed for the low $U$ is in agreement with previous 	
studies~\cite{PhysRevLett.94.056404,PhysRevLett.102.156402}. 
Figure~\ref{Fig2:tdos_ados_center}(c) depicts the $W_c^U$ as a function of $1/N_c$ 
at $U=0.0$ and $U=0.2$ for the half-filled AHM. Note the systematic and fast
convergence of W$_c$ with $N_c$ for both cases.

We further show in Fig.~\ref{Fig2:pseudo_gap} the evolution of the TDoS($\omega$) for $N_c=12$ at $U=0.0$ 
(Fig.~\ref{Fig2:pseudo_gap}(a)) and 0.2 (Fig.~\ref{Fig2:pseudo_gap}(b)) for various $\delta=W/W_c^U$. For 
finite $U$ a soft-pseudogap, which is linear in $\omega$ (cf. Fig.\ref{Fig2:pseudo_gap}(d)) develops at
the Fermi energy (note, this is true irrespective of electron filling) at intermediate disorder strengths 
immediately before the system becomes localized. In Fig.~\ref{Fig2:pseudo_gap}(c), we show that the pseudogap 
is robust as a function of $U<1$.  Noting that we have only short-range interaction, this soft-pseudogap 
cannot be attributed to excitonic effects (which are negligible here) as in the Efros-Shklovskii 
theory~\cite{Efros1975}. Also, since it occurs only in the TDoS and even for $N_c=1$, it cannot be 
due to the multivalley structure of the energy landscape~\cite{Hiroshi2009} since a single-site cannot 
generate a multivalley energy landscape to sustain sets of metastable states, and it should be 
contrasted from the Altshuler-Aronov zero-bias anomaly, which is due to weak nonlocal interactions 
and weak disorder~\cite{Altshuler1979}. We ascribed this soft pseudogap to the same scenario, which 
causes well-defined mobility edge to only exist when $\mu \geqq \omega_\epsilon$. $U$ suppresses 
localization and increases the TDoS.  However, near the Fermi energy the phase space for scattering 
by $U$ is drastically reduced leading to the opening of a soft pseudogap. Put differently, the pseudogap 
is due to the suppression of inelastic scattering by $U$ due to the Pauli principle and energy conservation. 
It is linear (cf. Fig.\ref{Fig2:pseudo_gap}(d)), rather than quadratic in $\omega$, due to the lack of 
momentum conservation.

\textit{Conclusions}--	
Based on experiment, theory, and simulations, there is a growing consensus that the local density of 
states in a disordered system develops a highly skewed~\cite{PhysRevLett.106.163902}, log-normal 	
distribution~\cite{Anthony2010,PhysRevLett.111.066601,Wei2013} with a typical value given by the geometric	
mean that vanishes at the localization transition, and hence, acts as an order parameter for the ALT.  
New mean field theories for localization, including the TMT and its cluster extension, the TMDCA, 	
have been proposed. In this letter, we extend the TMDCA to weakly interacting systems using second order
perturbation theory. We find that weak local interactions lead to an increase in W$_c$, with the 
localization edge preserved when the chemical potential is at or above the mobility edge energy. 
For finite $U$ we observe a soft-pseudogap for values of the disorder strength just above $W_c^U$.
% in tandem with experiment.

%=======================================================================================================	
\textit{Acknowledgments}--	
This work is supported by NSF DMR-1237565 and NSF EPSCoR Cooperative Agreement EPS-1003897. 
Supercomputer support is  
provided by the Louisiana Optical Network Initiative (LONI) and HPC@LSU.% computing resources.
%=======================================================================================================	

%\bibliography{TMDCA_SOPT}	

\begin{thebibliography}{10}

\bibitem{Anderson}
P.~W. Anderson, Phys. Rev. {\bf 109}, 1492 (1958).

\bibitem{gang4}
E.~Abrahams, P.~W. Anderson, D.~C. Licciardello, and T.~V. Ramakrishnan, Phys.
  Rev. Lett. {\bf 42}, 673 (1979).

\bibitem{Wegner1975}
F.~Wegner, Z. Phys. B {\bf 35}, 207 (1979).

\bibitem{Kramer}
B.~Kramer and A.~MacKinnon, Rep. Prog. Phys. {\bf 56}(12), 1469 (1993).

\bibitem{Mirlin-RMP}
F.~Evers and A.~D. Mirlin, Rev. Mod. Phys. {\bf 80}, 1355 (2008).

\bibitem{Mott}
N.~Mott, Adv. Phys. {\bf 16}(61), 49 (1967);
\textit{ibid.} Proc. Phys. Soc., Sect. A {\bf 62}, 416 (1949);
\textit{ibid.} {\em Metal-Insulator Transitions, 2nd ed.\/} (Taylor and Francis,
  London, 1990).

\bibitem{50years}
E.~Abrahams (ed.) {\em 50 Years of Anderson Localization\/} (World Scientific,
  2010).

\bibitem{Belitz-Kirkpatrick-1994}
D.~Belitz and T.~R. Kirkpatrick, Rev. Mod. Phys. {\bf 66}, 261 (1994).

\bibitem{Anthony2010}
A.~Richardella, P.~Roushan, S.~Mack, B.~Zhou, D.~A. Huse, D.~D. Awschalom, and
  A.~Yazdani, science {\bf 327}, 665 (2010).

\bibitem{Kravchenko-2d}
S.~V. Kravchenko, G.~V. Kravchenko, J.~E. Furneaux, V.~M. Pudalov, and
  M.~D'Iorio, Phys. Rev. B {\bf 50}, 8039--8042 (1994).

\bibitem{PhysRevB.76.165128}
K.~Maiti, R.~S. Singh, and V.~R.~R. Medicherla, Phys. Rev. B {\bf 76}, 165128
  (2007).

\bibitem{PhysRevB.71.125104}
K.~W. Kim, J.~S. Lee, T.~W. Noh, S.~R. Lee, and K.~Char, Phys. Rev. B {\bf 71},
  125104 (2005).

\bibitem{Sahu2011523}
R.~K. Sahu, S.~K. Pandey, and L.~Pathak, J. Solid State Chem. {\bf 184}(3), 523
  -- 530 (2011).

\bibitem{PhysRevB.74.104419}
A.~S. Sefat, J.~E. Greedan, G.~M. Luke, M.~Ni\'{e}wczas, J.~D. Garrett,
  H.~Dabkowska, and A.~Dabkowski, Phys. Rev. B {\bf 74}, 104419 (2006).

\bibitem{Raychaudhuri}
A.~Raychaudhuri, Adv. Phys. {\bf 44}(1), 21 (1995).

\bibitem{sanchez2010disordered}
L.~Sanchez-Palencia and M.~Lewenstein, Nature Physics {\bf 6}(2), 87
  (2010).

\bibitem{1751-8121-45-14-143001}
B.~Shapiro, J. Phys. A: Math. and Theor. {\bf 45}(14), 143001 (2012).

\bibitem{PhysRevLett.102.055301}
M.~White, M.~Pasienski, D.~McKay, S.~Q. Zhou, D.~Ceperley, and B.~DeMarco,
  Phys. Rev. Lett. {\bf 102}, 055301 (2009).

\bibitem{Kondov13}
S.~S. Kondov, W.~R. McGehee, W.~Xu, and B.~DeMarco, Phys. Rev. Lett. {\bf 114},
  083002 (2015).

\bibitem{Altshuler}
B.~L. Altshuler and A.~G. Aronov, JETP Lett. {\bf 27}, 662 (1978).

\bibitem{Efros1975}
A.~L. Efros and B.~I. Shklovskii, J. Phys. C: Solid State Physics {\bf 8}(4),
  L49 (1975).

\bibitem{Finkel1983}
A.~Finkel'stein, JETP {\bf 57}, 97 (1983).

\bibitem{Castellani1984}
C.~Castellani, C.~Di~Castro, P.~A. Lee, and M.~Ma, Phys. Rev. B {\bf 30},
  527--543 (1984).

\bibitem{Punnoose14102005}
A.~Punnoose and A.~M. Finkel'stein, Science {\bf 310}(5746), 289 (2005).

\bibitem{Anissimova}
S.~Anissimova, S.~V. Kravchenko, A.~Punnoose, A.~M. Finkel/'stein, and T.~M.
  Klapwijk, Nat Phys {\bf 3}(10), 707--710, ISSN 1745-2473 (2007).

\bibitem{PhysRevB.89.081107}
C.~E. Ekuma, H.~Terletska, K.-M. Tam, Z.-Y. Meng, J.~Moreno, and M.~Jarrell,
  Phys. Rev. B {\bf 89}, 081107 (2014).

\bibitem{Vlad2003}
V.~Dobrosavljevi\'{c}, A.~A. Pastor, and B.~K. Nikoli\'{c}, EPL {\bf 62}(1), 76
  (2003).

\bibitem{Vollhardt_PDF}
G.~Schubert, J.~Schleede, K.~Byczuk, H.~Fehske, and D.~Vollhardt. 
Phys. Rev. B {\bf 81}, 155106 (2010).

\bibitem{Wei2013}
W.~Li, X.~Chen, L.~Wang, Y.~He, Z.~Wu, Y.~Cai, M.~Zhang, Y.~Wang, Y.~Han, R.~W.
  Lortz, Z.-Q. Zhang, P.~Sheng, and N.~Wang, Sci. Rep. {\bf 3}, 1772 (2013).

\bibitem{PhysRevLett.110.066401}
M.~C.~O. Aguiar and V.~Dobrosavljevi\ifmmode~\acute{c}\else \'{c}\fi{}, Phys.
  Rev. Lett. {\bf 110}, 066401 (2013).

\bibitem{PhysRevLett.102.156402}
M.~C.~O. Aguiar, V.~Dobrosavljevi\ifmmode~\acute{c}\else \'{c}\fi{},
  E.~Abrahams, and G.~Kotliar, Phys. Rev. Lett. {\bf 102}, 156402 (2009).

\bibitem{Vollhardt}
K.~Byczuk, W.~Hofstetter, and D.~Vollhardt, Int. J. Mod. Phys. B {\bf 24}, 1727
  (2010).

\bibitem{PhysRevB.72.035122}
A.~N. Rubtsov, V.~V. Savkin, and A.~I. Lichtenstein, Phys. Rev. B {\bf 72},
  035122 (2005).

\bibitem{Rubtsov2004}
A.~N. Rubtsov and A.~I. Lichtenstein, JETP Lett. {\bf 80}, 61 (2004).

\bibitem{RevModPhys.83.349}
E.~Gull, A.~J. Millis, A.~I. Lichtenstein, A.~N. Rubtsov, M.~Troyer, and
  P.~Werner, Rev. Mod. Phys. {\bf 83}, 349 (2011).

\bibitem{0295-5075-82-5-57003}
E.~Gull, P.~Werner, O.~Parcollet, and M.~Troyer, EPL {\bf
  82}(5), 57003 (2008).

\bibitem{Fuchs2011}
S.~Fuchs, E.~Gull, M.~Troyer, M.~Jarrell, and T.~Pruschke, Phys. Rev. B {\bf
  83}, 235113 (2011).

\bibitem{PhysRevLett.56.2521}
J.~E. Hirsch and R.~M. Fye, Phys. Rev. Lett. {\bf 56}, 2521--2524 (1986).

\bibitem{PhysRevB.76.205120}
N.~Bl\"umer, Phys. Rev. B {\bf 76}, 205120 (2007).

\bibitem{Gebhard2003}
F.~Gebhard, E.~Jeckelmann, S.~Mahlert, S.~Nishimoto, and R.~Noack, Euro. Phys.
  B {\bf 36}(4), 491--509 (2003).

\bibitem{681704}
M.~Frigo and S.~Johnson.
\newblock ``FFTW: An adaptive software architecture for the FFT,'' 
” In {\em Proc. IEEE Int'l Conf. Acoustics, Speech, and Signal 
Processing}, vol.~3, Seattle, WA, 1381, (1998).

\bibitem{TMDCA-SOPT_Supp}
See Supplemental Material at [URL will be inserted by publisher] for the
  benchmarking of our developed TMDCA-SOPT with quantum Monte-Carlo simulations
  both for single-site (N$_c=1$) and finite cluster (N$_c=14$) for the
  Anderson-Hubbard model. As it is evident from the
  comparisons, the agreement in the parameter regime we explored is remarkable
  even up to relatively large value of $U$. This good agreement not only
  validates our assumption that higher order diagrams are suppressed by at
  least $\sim(U^3)$, but provides a comprehensive benchmark of the
  capabilities of the developed finite cluster formalism.

\bibitem{Ma-1982}
M.~Ma, Phys. Rev. B {\bf 26}, 5097 (1982).

\bibitem{PhysRevLett.91.066603}
D.~Tanaskovi\ifmmode~\acute{c}\else \'{c}\fi{},
  V.~Dobrosavljevi\ifmmode~\acute{c}\else \'{c}\fi{}, E.~Abrahams, and
  G.~Kotliar, Phys. Rev. Lett. {\bf 91}, 066603 (2003).

\bibitem{Golubev1998164}
D.~Golubev and A.~Zaikin.
\newblock Physica B {\bf 255}, 164 (1998); 
\textit{ibid.} D.~S. Golubev, C.~P. Herrero, and A.~D. Zaikin. 
\newblock EPL {\bf 63}, 426 (2003).

\bibitem{PhysRevLett.94.056404}
K.~Byczuk, W.~Hofstetter, and D.~Vollhardt, Phys. Rev. Lett. {\bf 94}, 056404
  (2005).

\bibitem{Hiroshi2009}
H.~Shinaoka1 and M.~Imada, J. Phys. Soc. Jpn. {\bf 78}, 094708 (2009).

\bibitem{Altshuler1979}
B.~L. Altshuler and A.~G. Aronov, Sov. Phys. JETP {\bf 50}, 968 (1979); 
\textit{ibid.} Solid State Communs. {\bf 30}, 
115 (1979).

\bibitem{PhysRevLett.106.163902}
R.~Sapienza, P.~Bondareff, R.~Pierrat, B.~Habert, R.~Carminati, and N.~F. van
  Hulst, Phys. Rev. Lett. {\bf 106}, 163902 (2011).

\bibitem{PhysRevLett.111.066601}
I.~S. Burmistrov, I.~V. Gornyi, and A.~D. Mirlin, Phys. Rev. Lett. {\bf 111},
  066601 (2013).

\end{thebibliography}
	
\end{document}

\title{Metal-Insulator-Transition in a Weakly interacting Disordered Electron System: Supplementary Notes} 

%Authors
\author{C. E. Ekuma}
\altaffiliation{Electronic address: cekuma1@lsu.edu}
\affiliation{Department of Physics \& Astronomy, Louisiana State University,
Baton Rouge, Louisiana 70803, USA}
\affiliation{Center for Computation and Technology, Louisiana State University, Baton Rouge, Louisiana 70803, USA}

\author{S.-X. Yang}
\affiliation{Department of Physics \& Astronomy, Louisiana State University,
Baton Rouge, Louisiana 70803, USA}
\affiliation{Center for Computation and Technology, Louisiana State University, Baton Rouge, Louisiana 70803, USA}

\author{H. Terletska}
\affiliation{Department of Physics \& Astronomy, Louisiana State University,
Baton Rouge, Louisiana 70803, USA}
\affiliation{Center for Computation and Technology, Louisiana State University, Baton Rouge, Louisiana 70803, USA}

\author{K.-M. Tam}
\affiliation{Department of Physics \& Astronomy, Louisiana State University,
Baton Rouge, Louisiana 70803, USA}
\affiliation{Center for Computation and Technology, Louisiana State University, Baton Rouge, Louisiana 70803, USA}

\author{N. S. Vidhyadhiraja}
\affiliation{Theoretical Sciences Unit, Jawaharlal Nehru Center for Advanced Scientific Research, Bangalore, 560064, India}

\author{J. Moreno}
\affiliation{Department of Physics \& Astronomy, Louisiana State University,
Baton Rouge, Louisiana 70803, USA}
\affiliation{Center for Computation and Technology, Louisiana State University, Baton Rouge, Louisiana 70803, USA}

\author{M. Jarrell}
\altaffiliation{Electronic address: jarrellphysics@gmail.com}
\affiliation{Department of Physics \& Astronomy, Louisiana State University,
Baton Rouge, Louisiana 70803, USA}
\affiliation{Center for Computation and Technology, Louisiana State University, Baton Rouge, Louisiana 70803, USA}
%=======================================================================================================    

\pacs{72.15.Rn,02.70.Uu,64.70.Tg,71.23.An,71.27.+a}
\maketitle

\section{Details of the Formalism}
In our formulation of both the dynamical cluster approximation (DCA) and typical medium DCA (TMDCA), 
we treat both disorder and interactions using the Anderson-Hubbard Hamiltonian
\begin{equation}	
  \label{eqn:model}	
H=-\sum_{\langle i j \rangle \sigma}t_{ij}(c_{i \sigma}^{\dagger}c^{\phantom{\dagger}}_{j \sigma}+h.c.)+\sum_{i \sigma} (V_i-\mu)n_{i \sigma} + 	
U\sum_{i} n_{i \uparrow} n_{i \downarrow}.	
\end{equation}	
where symbols have their usual meanings are described in the main text.
For the TMDCA, instead of using the conventional Matsubara frequency approach, we 
reformulate our formalism in real frequency. 
This facilitates the analysis of the zero-temperature physics, where Matsubara-frequency-based approaches 
may not be adequate. It also avoids the 
difficulty of analytical continuation of the major observables, the Green function, and the self-energy, 
from the Matsubara-frequency to real frequency. 
To achieve this, we recall that the second order perturbation theory self-energy is the expansion up 
to the second order in $U$ (\orderof$\left[ U^2\right]$) around the Hartree-corrected host propagator given as
\begin{equation}
 \Sigma^{int}_c(i,j,\omega) = \Sigma^{H}_c[{\cal \tilde{G}}]+\Sigma^{(SOPT)}_c[{\cal \tilde{G}}]
\end{equation}
where $i$ and $j$ are site indices. The first 
term $\Sigma^{H}_c[{\cal \tilde{G}}]=U\tilde{n_i}/2$ is the  
Hartree term, while the second term is the density-density term which in the Matsubara frequency is
\begin{widetext}
\begin{equation}\label{eqn:Matsubara}
\Sigma^{(SOPT)}_c[{\cal \tilde{G}}]
=
-\lim_{i\omega\rightarrow \omega^+} \left[ \frac{U^2}{N_c^2\beta^2}\sum_{m,p,\P,\Q} {\cal \tilde{G}}(\K+\Q,i\omega_n+i\nu_m) 
 {\cal \tilde{G}}(\P+\Q,i\omega_p+i\nu_m) {\cal \tilde{G}}(\P,i\omega_p) \right]
\end{equation}
\end{widetext}
where $\beta$ is the inverse temperature. We note further that the analytic continuation process may miss 
important features of the observable being studied if not done carefully aside its inability to study 
zero-temperature physics, which coincidentally, is the regime we are interested in the present study.
Thus, to avoid such analytic continuation, and since our cluster problem is solved in real space, more also, 
for the fact that it is numerically more advantageous to work in the real 
frequency than in the Matsubara frequency, we convert 
the Matsubara sums (Eq.~\ref{eqn:Matsubara}) to real frequency integrals using the spectral representation: 
$ {\cal \tilde{G}}(i\omega)=\int d \epsilon \displaystyle \rho({\cal \tilde{G}}(\epsilon))/(i\omega - \epsilon)$, 
where $\rho({\cal \tilde{G}}(\epsilon))=-\frac{1}{\pi}\Im {\cal \tilde{G}}(\epsilon)$. 
Defining $\rho_{\Sigma} (i,j,\omega) = -\frac{1}{\pi} \Im \Sigma^{(SOPT)} (i,j,\omega)$, the second 
order term of the self-energy in real frequency is

\begin{widetext}
\begin{equation}\label{eqn:spectrarep}
\rho_{\Sigma} (\omega)
=
U^2 \int d\epsilon_1 d\epsilon_2 \rho_{ \cal \tilde{G} }(\epsilon_1) \rho_{ \cal \tilde{G} }(\omega - \epsilon_1 +
\epsilon_2) \rho_{ \cal \tilde{G} }(\epsilon_2) \times \left[n_f(-\epsilon_1) n_f(\epsilon_2) n_f(-\omega+\epsilon_1-\epsilon_2) 
 + n_f(\epsilon_1) n_f(-\epsilon_2) n_f(\omega-\epsilon_1+\epsilon_2)\right]
\end{equation}
\end{widetext}
where site labels have been suppressed, 
$\rho_{ \cal \tilde{G} }$ = $-1/\pi \Im { \cal \tilde{G} }$, and $n_f = 1/(e^{\beta\epsilon}+1)$ is the Fermi 
function. We note that $\rho_{ \cal \tilde{G} }$ vanishes only for $| \omega | \geq 3B/2$~\cite{Gebhard2003}, 
where $B$ is the full bandwidth ($12t =3$ in our unit). The real part of the second order term of the 
interacting self-energy, $\Sigma^{SOPT}_R(i,j,\omega$) on each cluster site is obtained via the Hilbert transform 

\begin{equation}
 \Sigma^{SOPT}_R(i,j,\omega)=
\int d \omega' \displaystyle \frac{\rho_{\Sigma} (i,j,\omega')}{\omega - \omega'}.
\end{equation} 
Since $1/(x+i 0^+) = \specialp(1/x)-i\pi \delta(x)$ (where `$\specialp$' is the principal value of the integral), 
the non-local SOPT self-energy is then
\begin{equation}
 \Sigma_c^{SOPT}(i,j,\omega) = \Sigma^{SOPT}_R(i,j,\omega) - i\pi \rho_{\Sigma} (i,j,\omega).
 \end{equation}
We note that Eq.~\ref{eqn:spectrarep} 
scales as an \orderof$\left[ N^3\right]$, where $N$ is the number of the grid points used for the integration. This $N^3$ process is 
dramatically reduced to scale logarithmically as $N \ln N$ using fast Fourier transformation
~\cite{681704}. 

\section{Benchmarking the TMDCA-SOPT}
To validate our developed method and benchmark its suitability for studying the Anderson-Hubbard model, 
we compare our results for the Anderson-Hubbard model at half-filling with results 
from the dynamical cluster approximation (DCA) using the continuous-time quantum Monte Carlo method 
(CTQMC) ~\cite{PhysRevB.72.035122,Rubtsov2004,
RevModPhys.83.349,0295-5075-82-5-57003,Fuchs2011} as the cluster solver. 
The quantum Monte Carlo (QMC) methods are powerful tools that enable controlled calculations of the properties of large
quantum many-particle systems. Details of the CTQMC formalisms are well described in the literature 
(see for e.g., Refs.~\onlinecite{PhysRevB.72.035122,Rubtsov2004,RevModPhys.83.349,0295-5075-82-5-57003,Fuchs2011,
PhysRevLett.56.2521,PhysRevB.76.205120}) 
as such, and we will not attempt to give a detailed description of the algorithms here but will 
just give an overview which enables us to compare the QMC results with our Hilbert transformed 
imaginary frequency data. In the CTQMC, to avoid any possible 
time discretization error in the imaginary time axis, we adopt the recent improvements in the QMC algorithms 
in the continuous imaginary time~\cite{PhysRevB.72.035122,Rubtsov2004,0295-5075-82-5-57003,PhysRevB.76.205120}. 
This significantly improves the quality of the data. 

To make the comparison, we convert our real frequency data to Matsubara frequency using the Hilbert transformation, 
and obtain the local Green function and self-energy, respectively, as

\begin{subequations}
\begin{align} 
  G_{loc}(i\omega_n)= -\frac{1}{\pi}\int_{-\infty}^{\infty} d \omega \displaystyle \frac{\Im G_{loc}(\omega)}{i\omega_n-\omega}, \label{eqn:Glocaltransf} \\[1.8em]
  \Sigma_{loc}(i\omega_n)= -\frac{1}{\pi}\int_{-\infty}^{\infty} d \omega \displaystyle \frac{\Im \Sigma_{loc}(\omega)}{i\omega_n-\omega} \label{eqn:Sigmalocal}.
\end{align} 
\end{subequations}
where $\omega_n=(2n+1)\pi/\beta$ with $\beta=1/k_BT$ and $n\in\mathbb{Z}$ is the fermionic Matsubara frequency, 
and $k_B$ (=1) is the universal Boltzmann's constant.

\subsection{Limit of Zero Disorder: Hubbard Model}

\begin{figure}[htb]
%FIG. 1
 \includegraphics[trim = 0mm 0mm 0mm 0mm,width=1.0\columnwidth,clip=true]{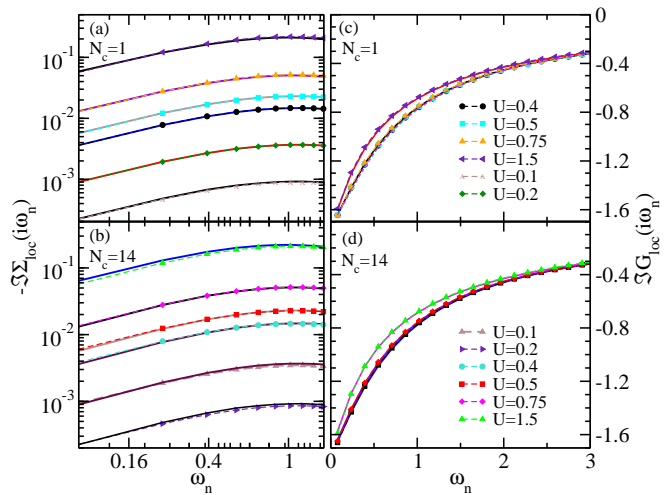}
\caption{(Color online). 
The comparison of the single-particle quantities from TMDCA-SOPT and DCA-CTQMC at $T=0.025 $ 
for the half-filled Hubbard model (zero disorder and the finite Coulomb interaction $U>0$).
The comparison of the imaginary part of the local self-energy for $N_c=1$ (a) and $N_c=14$ (b), 
respectively,  on a log-log scale. Also shown is the corresponding comparison of 
the local Green function for $N_c=1$ (c) and $N_c=14$ (d), respectively on a linear scale. 
Smaller values ($U<0.3$ for $N_c=1$) and 
($U<0.4$ for $N_c=14$) are not shown on the local Green functions plot for easy readability as they are too 
close to each other. In both cases, the dotted lines with symbols are for the TMDCA-SOPT while the solid lines 
depict the DCA-CTQMC results. Further, for the self-energy plots, smaller interaction strengths ($U\leq0.5$) are on top of 
each other and as such, may appear indistinguishable.} 
\label{Fig:compare_CTQMC_1} 
\end{figure}

\begin{figure*}
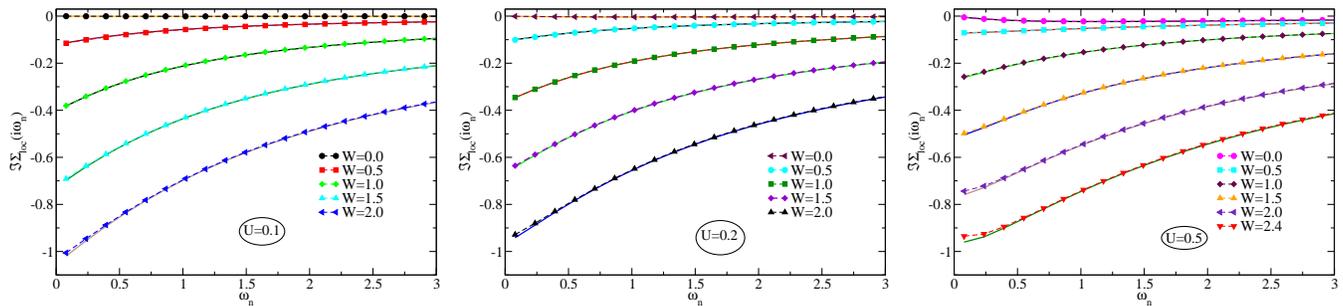

%FIG. 2
  \begin{center}
    \begin{minipage}[t]{0.32\linewidth}
      \raisebox{-0cm}{{\includegraphics[clip,angle=0,width=1.0\columnwidth,clip=true]{Slocal_TMDCA-SOPT_CTQMC_Nc14-U0.1W_b40.eps}}}
    \end{minipage}\hspace{0.0cm} 
\hfil
    \begin{minipage}[t]{0.32\linewidth}
      \raisebox{-0cm}{{\includegraphics[clip,angle=0,width=1.0\columnwidth,clip=true]{Slocal_TMDCA-SOPT_CTQMC_Nc14-U0.2W_b40.eps}}}
    \end{minipage}
\hfil
    \begin{minipage}[t]{0.32\linewidth}
      \raisebox{-0cm}{{\includegraphics[clip,angle=0,width=1.0\columnwidth,clip=true]{Slocal_TMCA-SOPT_CTQMC_Nc14-U0.5W_b40.eps}}}
    \end{minipage}
\caption{\label{Fig:compare_CTQMC_UW}(Color online). 
The comparison of the self-energy from TMDCA-SOPT and DCA-CTQMC at $T=0.025 $ 
for the half-filled Anderson-Hubbard model.
From left to right, the Coulomb interaction $U$ is fixed at 0.1, 0.2, and 0.5, respectively.
Here, the solid lines are for the DCA-CTQMC 
while the dash lines with closed symbols are for the TMDCA-SOPT. }  
\end{center}
\end{figure*}

As a natural consequence, the self-energy of the Hubbard model of the numerically exact DCA-CTQMC contains the most vital information 
for benchmarking with a mean-field theory like the TMDCA-SOPT. We show in Figs.~\ref{Fig:compare_CTQMC_1}(a) and 
~\ref{Fig:compare_CTQMC_1}(b) the comparison plots of our TMDCA-SOPT (using Eq.~\ref{eqn:Sigmalocal}) as compared with the 
DCA-CTQMC data for the $N_c=1$ and finite cluster, 14, respectively, at various values of the interaction. As it is evident 
from the plots, our data benchmarks well up to $U\sim2.25$ for $N_c=1$ and $U\sim0.75$ for $N_c=14$ showing 
both quantitative and qualitative agreement between our TMDCA-SOPT data and the DCA-CTQMC results. The remarkable 
exact agreement between the TMDCA-SOPT and DCA-CTQMC for low U ($\leq0.75$), at all frequencies for both single-site and finite 
cluster shows that our self-energy has the correct behavior and as such, ensures that the conclusions arrived in the main 
paper are numerically correct. At least for the smaller $U$-values (which is the regime we are interested in), 
the good agreement with DCA-CTQMC further shows that the perturbation expansion of the self-energy up to 
\orderof$\left[ (U/B)^2\right]$ in $U$ should capture all the dominant quantum fluctuations 
guaranteeing that quantitatively correct results are obtained. Hence, the results we 
presented in the main paper which are for small interacting disordered electron system are accurate to 
within the computational accuracy of our formalism.  

We further show in Figs.~\ref{Fig:compare_CTQMC_1}(c) and ~\ref{Fig:compare_CTQMC_1}(d), the plot of the imaginary local Green function of our data 
(transformed from real to Matsubara frequency using Eq.~\ref{eqn:Glocaltransf}) as compared to the DCA-CTQMC data 
for $T=0.025 $ and $W=0$ for the $N_c=1$ and $N_c=14$, respectively. Again, as a confirmation of the good agreement in the local 
self-energy, the local 
Green function from the two methods is numerically the same especially for the small $U$-values. They 
are practically on top of each other up to $U\approx1.50$ for both $N_c=1$ and 14.

\subsection{Finite Disorder and Interaction: Anderson-Hubbard Model}
To further check the applicability of the expansion of the interacting 
self-energy in powers of $U$ up to the second order for 
the study of interacting disordered electron system, we further benchmark our developed method 
for the Anderson-Hubbard model with the DCA-CTQMC. This again becomes imperative as we are not 
aware of a prior benchmarking especially for the finite cluster.

We show in Figs.~\ref{Fig:compare_CTQMC_UW} the comparison of our developed 
method for fixed interaction strengths ($U=0.1$, 0.2, and 0.5, respectively) at various disorder strengths, respectively 
(using Eq.\ref{eqn:Sigmalocal}), with the 
imaginary frequency data of DCA-CTQMC. As it is evident from the plots, there is almost a perfect 
agreement between our Hilbert transformed data with the DCA-CTQMC results 
even for relatively large value of the interaction strengths ($U\sim0.5$) for all the disorder strengths up 
to the localization transitions. 
This further confirms that our truncation of the perturbation series expansion of 
the interacting self-energy at \orderof$\left[ (U/B)^2\right]$ (at least in the weak interaction regime) is enough to account 
for the quantum fluctuations in the typical environment. One can thus affirm that at least, within the weak 
interaction regime ($U/4t \ll 1$) in a disordered electron system that higher order diagrams in the expansion of 
the $\Sigma^{int}_c$ are suppressed by at least $\sim(U^3)$. 

\section{Exploring the Mobility Edge}
%=======================================================================================================    
\begin{figure}[b]    
%FIG. 3    
 \includegraphics[trim = 0mm 0mm 0mm 0mm,width=1\columnwidth,clip=true]{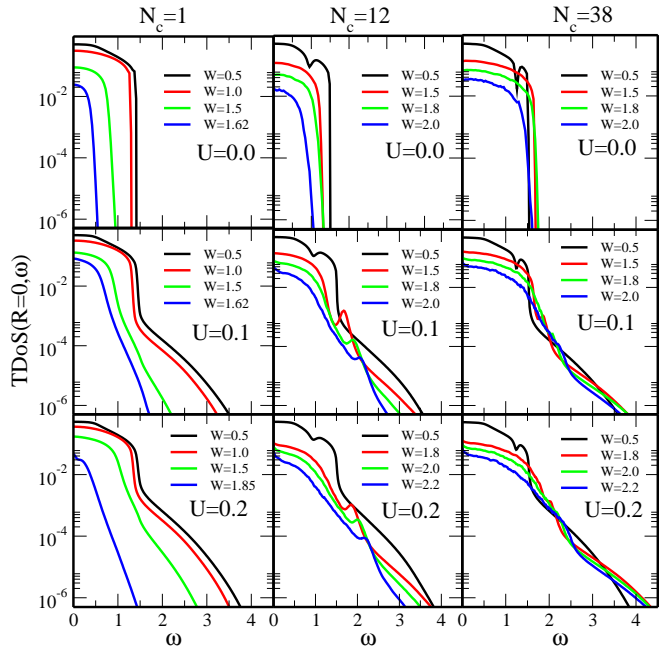}    
\caption{(Color online). The typical density of states of the Anderson-Hubbard model at half-filling for     
increasing U-value for the TMT ($N_c=1$) and finite clusters ($N_c=12$ and 38) at various disorder     
strengths on a log-linear plot. Observe that the $U=0$ results have 
infinite slopes signifying the localization edge while for all the interactions, 
even for a very small $U=0.1$ which is 1/30th the bandwidth, the slopes remain finite out to the band edge 
of the ADoS (see the main text)     
consistent with the mixing of the localized and extended states with different energies leading 
to the suppression of the mobility edge (see main text).}     
\label{Fig:TDoS_log-linear_ed0}      
\end{figure}    
%=======================================================================================================
We show in Fig.~\ref{Fig:TDoS_log-linear_ed0}, the plot of the typical density of states on 
a log-linear scale at half-filling for the TMT ($N_c=1$) and for the finite clusters ($N_c=12$ and 38) 
for increasing $U$-values at different disorder strengths for 3D. 
Clearly for $U=0$, the slopes of the TDoS for both the TMT ($N_c=1$) and  TMDCA ($N_c=12$ and 38)     
becomes infinity signifying the existence of a sharp, well-defined mobility edge. However, for a very small    
$U=0.1$ ($1/30$ of the bandwidth), and for both the TMT and TMDCA-SOPT, the slope is evidently     
finite up to high frequencies signifying that the sharp well-defined localization edge is replaced by 
an exponentially fast cross-over. Hence, the incorporation of Coulomb interactions in the presence of disorder 
at half-filling leads to long band tails that are exponentially decaying. 

One can argue that since the tails 
are far away from the chemical potential, $\mu$ when $\mu=0$, they do not generally participate in transport since 
only states in the proximity of the Fermi energy is excited. 
This can be attributed to the fact that at half-filling with finite $U$, 
interaction induces the mixing of the localized and extended states with different energies of the states at the 
mobility edge leading to the suppression of the sharp, well-defined mobility edge boundary with 
the emergence of long tails with exponentially fast crossover. This is consistent with the delocalization 
nature of $U$ at $\omega=0$ states at half-filling leading to the increase in $W_c$.
As demonstrated in the main text, the mobility edge is well defined even in the presence of 
interactions as long as the chemical potential is located at or beyond the 
mobility edge energy of $U=0$. Hence, Fig.~\ref{Fig:TDoS_log-linear_ed0} essentially shows 
that at half-filling, due to the mixing of states induced by the interaction, the mobility edge energies 
develop tails with exponentially fast crossover which masks its detection. However, if we move the chemical 
potential energy at or outside the mobility edge energy (see the main text), the sharp, well-defined mobility 
edge is restored. 

The masking of the localization edge in the TDoS due to the mixing of the the states induced    
by the small interaction can further be confirmed from the convolutions point of view. 
This can be seen from the convolutions found in the second order (and higher)    
diagrams. These convolutions will mix the states above and below 
the non-interacting localization edge. Consider two such states: one localized and the other extended,     
which are now degenerate due to this mixing. Since the elastic disorder scattering causes these 
states to hybridize with each other, both states will become extended~\cite{Mott}.

%\bibliography{../TMDCA_SOPT}